# PSEUDO-MAGNETIC FIELDS IN GRAPHENE IN EXCESS OF 300 T: THEORETICAL FRAMEWORK

VICTOR ATANASOV

*Department of Condensed Matter Physics*
*Faculty of Physics, St. Kliment Ohridski University of Sofia*

*Виктор Атанасов.* ПСЕВДОМАГНИТНИ ПОЛЕТА В ГРАФЕН НАД 300 T: ТЕОРЕТИЧЕН ПОДХОД

Експерименталното получаване на псевдомагнитни полета над 300 T в наномехурчета от графен [2] представлява сериозно предизвикателство за настоящата теория, свързваща появата на калибровъчни потенциали от опън в кристалната решетка. Тук предлагаме теоретичен подход, в рамките на който магнитното поле може да се пресметне по-точно. Основна характеристика на този подход е взимането под внимание на тримерната вълнова функция на носителите в графена, като динамиката е постепено ограничена върху повърхнината на графена. В резултат на това геометрично обусловено калибровъчно поле се появява в двумерното уравнение, описващо повърхнинната динамика. Магнитното поле, свързано с този калибровъчен геометрично породен потенциал, има стойнности, близки до експерименталните.

*Victor Atanasov.* PSEUDO-MAGNETIC FIELDS IN GRAPHENE IN EXCESS OF 300 T: THEORETICAL FRAMEWORK

The experimental demonstration of pseudo-magnetic fields exceeding 300 T in graphene [2] nanobubbles represents considerable challenge for the present theory connecting the emergence of gauge fields due to strain in the underlying lattice. Here we propose a theoretical framework within which the magnitude of the pseudo-magnetic fields can be computed more accurately. The basic feature of this framework is that the carriers in graphene are considered with their three dimensional wave function which is then gradually constrained to the graphene surface. In the process, a geometrically induced gauge field emerges in the two dimensional equation for the surface dynamics. The computation of the magnetic field associated with this gauge potential reproduces the measured field strength.

***Keywords:*** graphene, pseudo-magnetic field, curvature, reduced dimensionality
***PACS numbers:*** 71.10.Pm, 02.40.-k, 72.80.Vp

———————
*For contact*: Victor Atanasov, St. Kliment Ohridski University of Sofia, Faculty of Physics, Department of Condensed Matter Physics, 5 James Bourchier Blvd., Sofia 1164,
phone: 359 2  8161703, e-mail: vatanaso@gmail.com

1. INTRODUCTION

The unrivalled flexibility and strength of graphene membranes [1] coupled with the large strain-induced fields observed just recently [2] suggest that strain engineering is a viable method for controlling the electronic properties of graphene, even at room temperature. The experimental demonstration of pseudo-magnetic fields exceeding 300 T provides venue for the study of the electronic properties of condensed matter systems in extremely high magnetic fields.

Therefore, it is of utmost importance to have a theoretical framework for predicting these pseudo-magnetic fields in graphene.

A distortion of the graphene lattice creates large, nearly uniform pseudo-magnetic fields and gives rise to a pseudo-quantum Hall effect [3]. This effect is unique to graphene because of its massless Dirac fermion-like band structure and particular lattice symmetry ($C_{3v}$), see Fig. 1 and Fig. 2 for a quick notion of graphene's lattice and band structure.

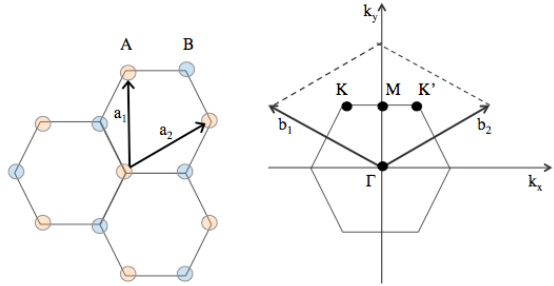

Fig.1. The lattice of graphene is comprised of two interpenetrating triangular lattices A and B with lattice unit vectors $a_1$ and $a_2$. Right: The Brillouin zone where the Dirac cones are located, $K$ and $K'$. The reciprocal lattice vectors are $b_1$ and $b_2$

Elastic strain is expected to induce a shift in the Dirac point energy from local changes in the electron density, as well as to induce an effective vector potential that arises from changes in the electron-hopping amplitude between carbon atoms, as well as the changes in the interatomic spacing. This strain-induced gauge field can give rise to large pseudo–magnetic fields ($B_s$) for appropriately selected geometries of the applied strain [6]. In this case the charge carriers in graphene are expected to circulate as if under the influence of an applied out-of-plane magnetic field. It has been proposed that a modest strain field with triangular symmetry will give approximately uniform magnetic field $B_s$ upward of tens of tesla [3].

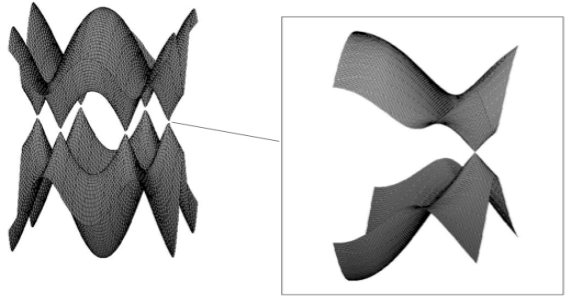

Fig. 2. Left: The electronic energy structure of graphene as a function of quasi-momentum taking values in the Brillouin zone[11]: $E_\pm = \pm t[3 + f(\mathbf{k})]^{1/2} - t' f(\mathbf{k})$, where $f(\mathbf{k}) = 2\cos(3^{1/2}k_y a) + 4\cos(3^{1/2}k_y a /2) \cos(3k_x a /2)$. Here $\mathbf{k} = (k_x, k_y)$, $t \approx 2{,}8$ eV, $t' \approx 0{,}1$ eV and $a = 1{,}42$ Å. The plus sign applies to the upper ($\pi$) and the minus sign to the lower ($\pi^*$) band. Right: The band structure zoomed in on one of the Dirac points, where $E_\pm = \pm \hbar v_F |\mathbf{k}|$. The Fermi velocity where $v_F$ is constant and the electron-hole symmetry is not broken if next-to-nearest neighbor hopping is neglected.

It is precisely this proposition which was tested in an experiment. Experimental spectroscopic measurements by scanning tunneling microscopy were obtained for highly strained nanobubbles that form when graphene is grown on a platinum (111) surface [2]. The nanobubbles exhibit Landau levels that form in the presence of strain-induced pseudo-magnetic fields greater than 350 T. Strained graphene nanobubbles were created by in situ growth of sub-monolayer graphene films in ultrahigh vacuum on a clean Pt(111) surface. Individual nanobubbles often have a triangular shape. This reflects the lattice symmetry of the graphene and the underlying Pt surface, and is typically 4 to 10 nm across and 0,3 to 2 nm tall. Atomic-resolution imaging of the nanobubbles confirms the honeycomb structure of graphene, although the lattice is distorted because of the large strain occurring in these structures [2].

Unfortunately, the underlying theory behind the measurements of these extreme pseudo-magnetic fields in graphene nanobubbles [2] produces field strengths which deviate considerably from the observed ones. This suggests a possible problem with the understanding of the mechanism creating these fields in graphene. The predicted pseudo-magnetic field, depicted on Fig.3, arising from strain field was calculated following [3].

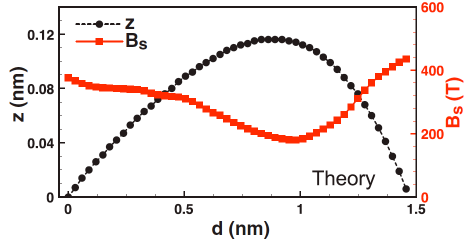

Fig.3. Theoretical estimation [2] of the pseudo-magnetic field on a simulated nanobubble. Notice the simulated bubble is twice as small (curvatures are twice as big) and produces substantially weaker fields. This figure appears as fig.3A in [2].

According to [3–5] a two dimensional strain field $u_{ij}(x,y)$ leads to a gauge field

$$\mathbf{A} = \begin{pmatrix} A_x \\ A_y \end{pmatrix} = \frac{\beta}{a} \begin{pmatrix} u_{xx} - u_{yy} \\ -2u_{xy} \end{pmatrix},$$

where $\beta = \partial \log(t)/\partial \log(a)$ relates the change in the hopping amplitude $t$ between nearest neighbour carbon atoms to bond length $a$; the $x$-axis is chosen along a zig-zag direction of the graphene latice. This yields the pseudo-magnetic field

$$B_s = \frac{\partial A_y}{\partial x} - \frac{\partial A_x}{\partial y}.$$

The large deviation, not only quantitatively but also qualitatively, from the measured field strength, as seen from Fig.4, points to a serious problem in the understanding of the mechanism creating the pseudo-magnetic fields in graphene.

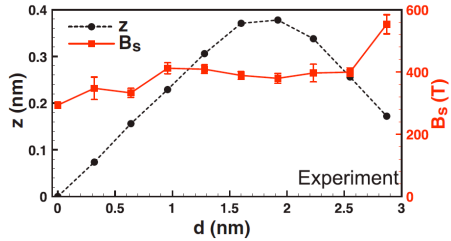

Fig.4. Experimental data is courtesy of [2]. This figure appears as fig.3A in [2].

Even more disturbing is the result from the magnetic field estimation using the heuristic relation

$$\Phi = \beta \frac{h^2}{la} \Phi_0$$

for the flux per ripple in distorted graphene sheet [6], where $\beta$ and $a$ appear in the previous formula, $h$ is the height, $l$ is the length and $\Phi_0$ is the quantum of flux. This formula applied for a nanobubble of $l = 4$ nm and $h = 0,5$ nm yields a $B_s$ of order 100 T. This result represents a deviation in excess of 250% from the measured field's strength! This by no means represents a trustworthy understanding of the mechanism creating the pseudomagnetic fields in graphene.

## 2. PARADIGM SHIFT

Here we argue that a paradigm shift in the way we envision carrier dynamics in graphene can produce the correct pseudo-magnetic fields, therefore a better understanding of the emerging property in graphene:

*global* topology/curvature → property → functionality relationship.

On one hand the starting point is the realization that graphene as a one-atom-thick membrane has carriers confined in a two dimensional space trapped in three dimensions. Furthermore the electrons are described by a massless relativistic equation. On the other hand the wavefunction of a quantum particle is always three dimensional due to the Heisenberg uncertainty principle which forbids setting any of the coordinates to zero (this would lead to indefiniteness in the momentum). In this way any two dimensional quantum motion would have an evanescent off-surface component of the wave function which can probe the two dimensional surface for curvature [7]. Indeed, the measured thickness of graphene, therefore extent of the evanescent off-surface component of the wavefunction, of graphene is almost 3 times as large, that is 0,34 nm, as the carbon-carbon distance of 0,14 nm, let alone the $p_z$-orbital. If the graphene sheet is curved, either intrinsically or extrinsically, then the carriers will at least be able to "feel" this curvature in the form of some effective mass or effective gauge field.

In fact, the curvature of the sheets builds strain and from a microscopic point of view strain has been shown to modify the electronic structure [3–5]. When interatomic distances are modified due to strain in the underlying lattice caused by curvature, the periodicity is disrupted and the conditions for the application of Bloch theorem do not hold.

Therefore, standard methods of solution fail to grasp the complexity of the rippling property of graphene. Effectively this means that one has to use a constraining procedure which starts from three dimensions and gradually confines the quantum dynamics onto the surface. Note, for the Schrödinger equation the curvature induces an effective (Da Costa) potential [7] through a confining procedure reducing the dimensionality of the quantum system. One question stands out: what is the corresponding potential for the relativistic case? We have answered this question and illustrated it through graphene in [8]. In short, the effective potential has a linear dependence on the Mean curvature $M$ as opposed to $M^2 - K$, where $K$ is the Gaussian curvature, in the usual (or nonrelativistic) case.

This effect is studied by expanding around the two dimensional space of the graphene sheet for vanishingly small excursions in the third direction. Such a procedure is well known for the Schrödinger equation [7] and we have carried it out in the case of a relativistic equation in a previous work [8]. The idea behind this confinement is that an "external force" gradually compresses the quantum dynamics onto a surface. This force can very well be the Coulomb electrostatic interaction, since as an electron leaves the surface of the sheet it disrupts the charge balance and an attractive "external force" appears confining the dynamics onto the surface. When the excursion in the normal to the surface direction becomes small enough one can take the limit (the normal to the surface coordinate goes to zero) and split the original 3+1 dimensional relativistic equation into 2+1 and 1+1 dimensional equations. The 2+1 dimensional equation encodes the quantum dynamics onto the surface and the 1+1 dimensional equation describes the behavior of the "evanescent component", that is the normal to the surface component of the wavefunction. In this way the Heisenberg principle is not violated since in the process described above no setting-to-zero of any coordinate of a three dimensional wavefunction takes place. According to a derivation [8]:

$$i\hbar v_F \gamma^0 \frac{\partial \chi_T}{\partial t} = -i\hbar v_F \sum_{a=1}^{2} \gamma^a D_a \chi_T - i\hbar v_F \gamma^3 M \chi_T,$$

$$i\hbar v_F \gamma^0 \frac{\partial \chi_N}{\partial t} = -i\hbar v_F \gamma^3 \frac{\partial \chi_N}{\partial q_3} + \left[V_\perp(q_3) + m(q_3)\right]\chi_N.$$

Here the $\gamma$-matrices belong to the fundamental 2-representation, $V_\perp(q_3)$ is the confining potential which depends on the normal to the surface coordinate $q_3$;

$D_a$ are the components of a two dimensional covariant derivative, $m(q_3)$ is a mass term and $v_F$ is the Fermi velocity in the material.

Note that the off-surface dynamics in this heuristic model can be solved in the nonrelativistic limit since as soon as the electrons leave the surface of the graphene sheet the symmetry which precludes the mass is no longer valid. The governing quantum dynamics is separated into surface $\chi_T$ and off-surface components $\chi_N$, where the off-surface "evanescent component" dynamics can be approximated with the wavefunction of a particle confined in a deep and narrow one dimensional potential well. For a detailed derivation and discussion refer to [8]. The magnitude of the geometric potential that comes from the confinement [8, 9] is

$$V = \hbar v_F M$$

and the vector potential that corresponds to this interaction is

$$A_\varphi = \frac{\hbar v_F M}{ec},$$

where $e = 1.6 \times 10^{-19}$ C the charge of the electron, $v_F \approx 110 \times 10^6$ m s$^{-1}$ is the Fermi velocity in graphene suspended by a substrate [10], $c=299792458$ m s$^{-1}$ is the speed of light, $\hbar=1{,}055.10^{-34}$ J s is the Planck's constant and $M = 0{,}5(\kappa_1 + \kappa_2)$ is the mean curvature, that is the average of the two principle curvatures $\kappa_{1,2}$ of the surface. From the relation $\mathbf{A} = 0{,}5 \mathbf{B} \times \mathbf{r}$ valid in two dimensions we obtain for the bump

$$B_s = \frac{2 A_\varphi}{r},$$

where $r = |\mathbf{r}| = r_1$ is the length of the radius vector from the axis of the bump to the periphery ($1/r_1$ is one of the principle curvatures of the bump) and $A_\varphi$ is the angular component of the vector potential in polar coordinates.

The computation for the magnetic field in our framework yields

$$B_s = \frac{2 \hbar v_F M}{ecr} = 241{,}9 \left( \frac{1}{r_1^2} + \frac{1}{r_1 r_2} \right),$$

where the radii of curvature of the bubble $r_{1,2}$ are given in [nm].

This formula applied for a nanobubble of $l = 4$ nm and $h = 0,5$ nm, that is $r_1 = 0,5l = 2$ nm and $r_2 = h$ yields a $B_s$ of the order of $B_s = 302$ T, which is the correct order of magnitude!

Unfortunately, the experimental data published in [2] is along one section of the graphene nanobubble. This is insufficient for determining the two radii of curvature of the bubble. Nevertheless, it is apparent that at the base of the bump, where it flattens out in a small expand one may expect small radii of curvature, that is large pseudo-magnetic fields. Indeed, the experiment shows magnetic field of the order of 550 T.

## 3. CONCLUSION

We have presented a simple, geometrically clear formula for a gauge potential capable of producing magnetic fields as observed in graphene nanobubbles. This gauge field emerges naturally in confining the quantum dynamics of graphene's electrons from the embedding space to the two dimensional surface. In this way one can compare different approaches towards explaining the origin of the pseudo-magnetic fields in graphene. In this paper, we have demonstrated a correct order of magnitude result, which is in favour of a paradigm shift in the perception of the long wavelength, low energy approximation to the relativistic dynamics of graphene's electrons.